\documentclass[aps,twocolumn,prl,showpacs]{revtex4}
\usepackage{graphicx}
\begin{document}


\title{Tunable Raman Photons from a Single Quantum Dot}



\author{G. Fernandez}\thanks{These authors contributed equally to this work.}
\author{T. Volz} \thanks{These authors contributed equally to this work.}
\author{R. Desbuquois}
\author{A. Badolato}
\author{A. Imamoglu}
\affiliation{Institute of Quantum Electronics, ETH Z\"urich,
Wolfgang-Pauli-Strasse 16, CH-8093 Z\"urich, Switzerland}


\date{\today}

\begin{abstract}
We report the observation of all-optically tunable Raman
fluorescence from a single quantum dot. The Raman photons are produced in an
optically-driven $\lambda$-system defined by subjecting the single electron
charged quantum dot to a magnetic field in Voigt geometry. Detuning the
driving laser from resonance, we tune the frequency of the Raman
photons by about 2.5~GHz. The number of scattered photons and the
linewdith of the Raman photons are investigated as a function of
detuning. The study presented here could form the basis of a new
technique for investigating spin-bath interactions in the solid
state.
\end{abstract}

\maketitle


Quantum dots (QDs) are often referred to as artificial atoms due to their
an-harmonic discrete energy level structure. Many key experiments with QDs such as single photon generation \cite{Michler,Santori}, spin pumping \cite{MeteScience,SteelVoigt,Gerardot} and coherent manipulation \cite{Gossard,VandersypenAC,Yamamoto} demonstrated their potential in quantum information processing \cite{LossQC}. However, as
compared to atoms and ions, QDs show a strong spectral
dispersion which is at present a serious limitation for their
usefulness, for example as sources of identical photons. While strong efforts
are undertaken to engineer QDs with identical spectral properties \cite{Schmidt}, applications in linear optical quantum computing or quantum repeaters appear out of reach with existing QD technology. One key element of a quantum repeater protocol is the creation of entanglement between two distant
quantum emitters by overlapping spectrally identical single photons
from two different quantum emitters on a beam splitter \cite{Cabrillo}. For ions,
such an entanglement scheme has been demonstrated recently in a
ground-breaking experiment by the Monroe group \cite{Monroe}. The realization
of a similar entanglement scheme based on solid-state emitters
remains an elusive goal up to date.\\
\indent In order to probabilistically entangle two QDs that are energetically close
an independent knob to ensure spectral overlap of the emitted photons is needed. In this
letter we demonstrate the first direct observation of
all-optically tunable spontaneous Raman fluorescence from a single
self-assembled QD. Using a cross-polarizer setup and a scanning
Fabry-Perot interferometer to suppress the excitation laser light,
we are able to detect the Raman-scattered photons on a CCD chip.
First, we demonstrate the frequency tunability of Raman photons with magnetic
field. While magnetic fields could in principle be used to tune the two emitters in
resonance with each other, all-optical manipulation is more
versatile, faster and allows spatial addressing of
several QDs within one sample. For this reason, in a second experiment the magnetic field is fixed and we vary the
excitation-laser frequency to optically tune the frequency of the
emitted photons over a range of roughly $2.5$~GHz. From the raw data, we extract the number of scattered photons, their center frequency and linewidth. As expected, the number of photons follows a Lorentzian as a function of detuning. Moreover, the data for the photon linewidths indicate a decrease down to the resolution limit of the Fabry-Perot as we tune off-resonance.\\
\indent Optical frequency tuning in a QD has been demonstrated
using the AC Stark effect in coherent QD spectroscopy
\cite{Wieck,SteelAC} and more recently using
resonantly scattered photons \cite{Mete,Shihresonant}. However, the
latter experiments are performed with laser powers far above QD
saturation. Not only does this pose additional technical challenges
to suppress the strong excitation laser, but also puts a lower bound
to the linewidth of the observed photons of 1.5 times the
spontaneous emission rate of the corresponding transition
\cite{Mete}. While resonantly scattered photons usually show at
least a lifetime limited spectral width, off-resonant Raman photons
can in principle be arbitrarily narrow and are only limited by the
laser linewidth and the low-energy spin coherence. For singly charged QDs as
studied here, the main decoherence mechanism of the metastable
ground state is the interaction of the electron spin with the
surrounding nuclear spins \cite{Khaetskii,Merkulov}. Hence, in
principle the spectral distribution of the Raman scattered photons
can give valuable information about this interaction and the work in this paper 
could form the basis of a new technique to study spin-bath
interactions in single QDs. We also note that resonantly scattered
Raman photons were previously used for spin read-out in a singly
charged QD \cite{Yamamoto}.\\
\indent The experiment is carried out with single self-assembled InGaAs QDs
that are charged with a single excess conduction-band electron and
that are subject to a magnetic field in Voigt geometry
\cite{SteelVoigt,KronerVoigt}. In Fig. 1 (a), the corresponding level
scheme is depicted. At zero magnetic field, ground and excited
(trion) states are degenerate. A finite magnetic field induces
Zeeman splittings according to the in-plane electron and hole
g-factors, $g_e$ and $g_h$. For large enough Zeeman splittings, each
of the trion states forms an independent $\lambda$-system together
with the two ground states. Optical selection rules determine the
polarization of both the vertical and diagonal transitions to be
linear but orthogonal to each other (denoted H and V in Fig. 1).
Experimentally this allows for efficient polarization separation of
the excitation laser from the photons of interest. The recombination
from the excited state takes place on a nanosecond timescale
with equal probability for decay into each ground
state. Note, that this is in stark contrast to the widely used
Faraday geometry, where this branching ratio is typically on the
order of $10^{-3}$ \cite{Jan}.

When the QD is resonantly driven on one of the optical transitions,
the electron spin is very efficiently pumped into the other ground
state and further photon
absorption or emission is stopped \cite{MeteScience,SteelVoigt}. Hence, for efficient photon
production by Raman scattering the spin state of the electron needs
to be restored on a short timescale. One possibility is to work at
the edges of the voltage range for which the QD is singly charged.
In this co-tunneling regime, the electron interacts with the
electrons of the Fermi sea in the back contact which leads to
spontaneous spin-flip events and effectively suppresses spin pumping
\cite{Jan}. Another possibility is the use of a second laser to optically
re-pump the electron spin \cite{MeteScience}. Both methods are applied in this
work.

In the experiment, two different QD samples with 25nm and 35nm
tunneling barriers between the QD layer and the n-doped back contact
are investigated. A transparent Schottky gate on the sample surface 
allows for determinist charging of the QD. The QD density is low (typically less than 0.1 per
$\mu$$m^{2}$) such that individual QDs can be studied using a
confocal microscopy setup with an almost diffraction-limited spot
size of 1~$\mu$$m^{2}$. The sample is immersed in a liquid-helium
bath cryostat at a temperature of 4.2~K, with magnetic fields up to 10~T. A stack of xyz
nano-positioners allows for precise positioning of the sample. QDs
are spatially and spectrally located by photoluminescence (PL)
spectroscopy. The emission wavelength of the QDs studied
is typically centered around 960nm. A differential transmission (DT) technique \cite{Benito,AlexDT} is used
to precisely determine the transition energies of the individual
QDs. The transmitted signal is split into its H and V polarization
components in order to separate vertical and diagonal transitions
from each other. The Raman-scattered photons are collected through
the confocal microscope and are sent through a fiber onto a CCD
camera for counting. The excitation laser light is suppressed by a
polarizer before the fiber. In addition, the collected light passes
through a Fabry-Perot interferometer before hitting the CCD
chip. This further suppresses the undesired background and serves
as a spectral filter for the collected photons. The interferometer
with parallel mirrors has a free spectral range of 15~GHz and
depending on the exact alignment a finesse of typically 40 to 60 is
achieved. The Fabry-Perot is stabilized using the transmission of an
independent titanium sapphire laser at 905nm. Both lasers are frequency-locked to a wavemeter
(High Finesse, WSU-30) which has an accuracy of better than 30~MHz.

\begin{figure}[tb!]
\begin{center}
\includegraphics[width=0.48 \textwidth]{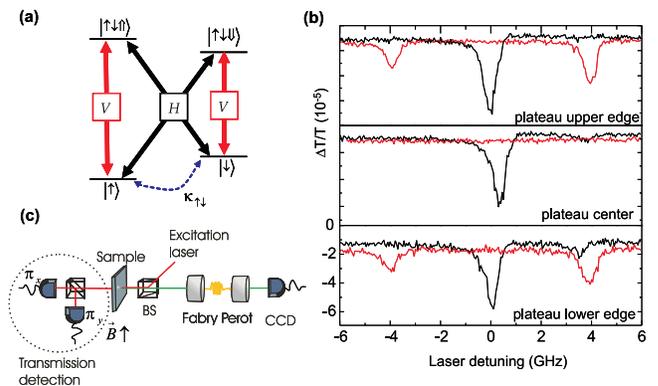}
\end{center}
\caption{(a) Level scheme of a singly charged quantum dot subject to a
magnetic field in Voigt geometry. The four transitions are linearly
polarized with the outer and inner transitions polarized
perpendicularly to each other (denoted H and V). (b)
Polarization-resolved differential transmission at $B=0.6~T$
for a quantum dot with $g_h \approx - g_e$. The inner transitions are degenerate and show no spin
pumping in contrast to the outer transitions. (c) Experimental set-up.
A Fabry-Perot interferometer suppresses the excitation laser light
and spectrally filters out the Raman photons. \label{Figure1}}
\end{figure}

\begin{figure}[tb!]
\begin{center}
\includegraphics[width=0.4 \textwidth]{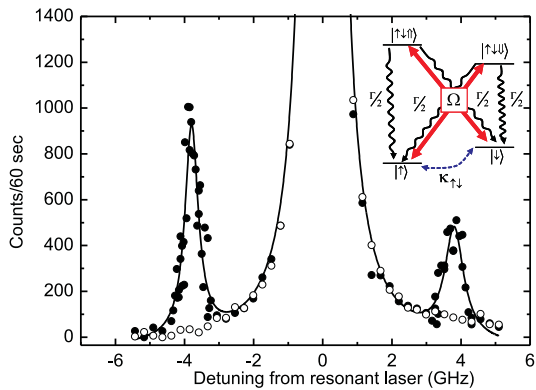}
\end{center}
\caption{Number of photons ($\bullet$) as a function
of the detuning from the excitation laser. Each data point is a single
shot with
an integration time of 60 seconds. The measurement is performed
at the upper edge of the plateau. The background ($\circ$)
was taken outside the charging plateau. The two peaks at -3.7~GHz and
3.9~GHz correspond to the two outer
transitions. The solid line is a guide to the eye.
\label{Figure2}}
\end{figure}

The
ground and excited state splittings are given by $g_e \mu_B B$ and $|g_h|\mu_B B$ respectively. We find
experimentally, that the electron g-factor $g_e$ is roughly the same
for all the QDs investigated here ($g_e \approx 0.45$), while the
in-plane hole g-factor $g_h$ shows a large distribution, in
accordance with previously reported results in literature \cite{KronerVoigt,Holes}.
  In particular, 2 out of 15 QDs had a hole
g-factor with approximately the same magnitude as the electron
g-factor but opposite sign, i.e. $g_h \approx - g_e$, which ensures
large enough splittings. Moreover, for this particular case the two diagonal
transitions are degenerate and consequently they exhibit no spin pumping, since the excitation laser acts both as pump and re-pump at
the same time. Figure \ref{Figure1}(c) demonstrates this effect: The
three polarization resolved DT traces were recorded at the two edges
and in the middle of the charging plateau for a magnetic field B$\approx$0.6~T. 
Whereas the outer transitions are efficiently spin-pumped
away from the plateau edges, the degenerate inner transitions
(central dip) exhibit no spin-pumping. We confirmed that spin
pumping is suppressed up to B$\approx$4~T.\\
\indent The efficient optical restoration of the electron spin makes this QD
a good candidate for observing resonantly scattered photons. Figure
\ref{Figure2} displays the number of detected photons after the
Fabry-Perot filter as a function of detuning from the driving laser
frequency which was set on the diagonal transition. The experiment
was performed again at a B$\approx$0.6~T. Even though the
laser light is suppressed by a polarizer and the Fabry-Perot filter,
it still gives rise to a strong central background peak. Besides
this central line, two distinct peaks are visible at -3.7~$\pm$0.2~GHz and +3.9~$\pm$0.2~GHz
 detuning, corresponding to photons emitted on the two
vertical transitions. Within our experimental resolution, the splitting is consistent with
the DT data from Figure~\ref{Figure1}(c) for which we obtained
a splitting of 7.9$\pm$0.1~GHz.  An obvious feature of the
data in Figure~\ref{Figure2} is the different height of the two
fluorescence peaks. Comparing with the DT data of
Figure~\ref{Figure1}(c), we note that the amplitude ratio for the
two transitions is approximately reversed. The origin of this asymmetry is not clear. 
The data presented in Figure~\ref{Figure2}
were taken for an excitation laser power at around saturation for
which the signal to noise ratio is optimal. 

\begin{figure}[tb!]
\begin{center}
\includegraphics[width=0.42 \textwidth]{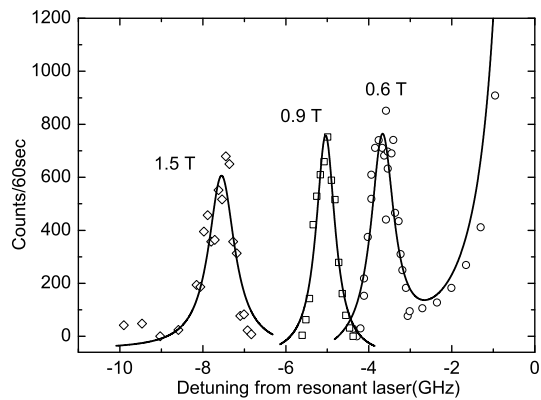}
\end{center}
\caption{Frequency tuning with magnetic field. The center frequency
of the Raman photons shifts with magnetic
field. Again, the solid lines are guides to the eye. \label{Figure3}}
\end{figure}

An important aspect in terms of quantum information processing is
the tunability of the resonantly scattered photons, with the magnetic
field being an obvious experimental knob. Figure~\ref{Figure3}
demonstrates this possibility. Here, the measurement of
Figure~\ref{Figure2} was repeated for three different magnetic
fields and the photons of the red-detuned outer transition were
recorded for a fixed laser frequency. Note, that this experiment was performed in the middle of
the charging plateau where the co-tunneling rates are negligible and the spin is optically re-pumped. The results demonstrate a
tunability range of roughly 4~GHz for a magnetic field variation of
0.9~T. The degeneracy of the inner transitions allows the frequency of the
driving laser to be kept constant for all three values of the
magnetic field.

Next, we turn our attention to off-resonant Raman scattering. At
typical powers used in the experiment, i.e. around saturation,
resonantly scattered Raman photons should have a linewidth
determined by dephasing of both excited and ground state and
potential inhomogeneous broadening due to e.g. charge fluctuations
in the environment of the QD \cite{Alex}. In contrast, for very low powers, we
would expect a narrowing of the fluorescence down to a linewidth
determined by the dephasing rates of the two ground states
$\gamma_{\uparrow\downarrow}$, which consists of the co-tunneling
rate to the back-contact, $\kappa_{\uparrow\downarrow}$, plus additional fluctuations due to
magnetic impurities in the enviroment. For large detunings, i.e. $\Delta \gg \Omega,
\Gamma, \gamma_{\uparrow\downarrow}$, the linewidth is solely
determined by $\gamma_{\uparrow\downarrow}$.
Hence, a photon source based on (far) off-resonant Raman scattering
could in principle produce photons that are limited by spin-dephasing.

\begin{figure}[t!]
\begin{center}
\includegraphics[width=0.49 \textwidth]{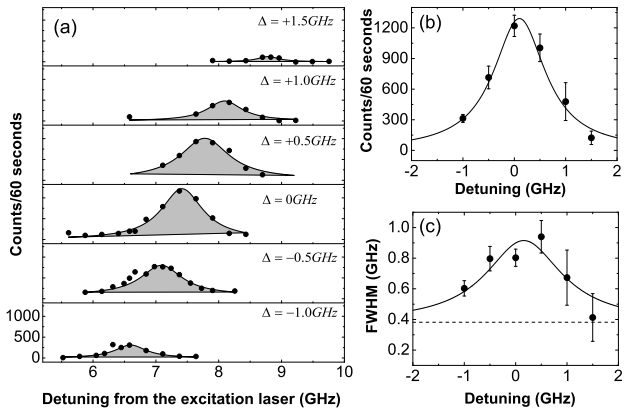}
\end{center}
\caption{(a) Optical tuning of Raman photons with the excitation laser 
detuning $\Delta$.(b) Overall number and (c) linewidth (FWHM) of Raman photons (both as a
function of $\Delta$). The dashed line represents the resolution limit of the Fabry-Perot. The solid line is a guide to the eye based on a Lorentzian fit with the resolution limit of the Fabry-Perot as the offset.\label{Figure4}}
\end{figure}

To demonstrate off-resonance Raman scattering we
study a QD in the 25~nm sample that has splittings of outer and
inner transitions of 10~GHz and 4.8~GHz, respectively, at B$\approx$1.2~T. 
The excitation laser frequency is tuned close to the
energetically higher outer transition. Measurements are performed in
the co-tunneling regime to ensure efficient restoration of the
spin-state by co-tunneling. Figure~\ref{Figure4}~(a) displays the
number of scattered Raman photons for different detunings $\Delta$
of the excitation laser from resonance. When the laser is detuned
from resonance (in steps of 0.5~GHz), the center frequency of the
scattered photons shifts accordingly and at the same time the number
of scattered photons decreases. Without changing the power of the
excitation laser a tuning range of about 2.5~GHz around resonance is
covered. In order to analyze the data quantitatively, we fit a
Lorentzian to each curve and extract center frequency, amplitude and
linewidth \cite{linewidth}. Results are displayed in Fig.~\ref{Figure4}(b) and (c).
We expect the number of Raman photons to follow a Lorentzian in
$\Delta$, i.e. to go as $1/\Delta^2$ for large detuning. The
Lorentzian fit in Fig.~\ref{Figure4}(b) has a width of $1.5$~GHz,
coinciding with the linewidth measured in differential transmission.

Figure ~\ref{Figure4}(c) demonstrates a decrease of the Raman photon
linewidth as we tune off resonance. From the on-resonance value of about
1~GHz, the observed linewidth decreases down to about
0.4~GHz at a detuning of 1.5~GHz.
The decrease in linewidth down to the 
resolution limit of the Fabry-Perot puts an upper bound 
to the dephasing rate $\gamma_{\uparrow\downarrow}$ and 
therefore also to the cotunneling rate
 $\kappa_{\uparrow\downarrow}$. Given a better frequency 
 resolution, the dephasing rate could be
mapped out as a function of the gate voltage by repeating the experiment in different regions 
of the plateau edge. Moreover, moving away from the plateau edges to the
middle of the plateau, we expect the cotunneling rate to become
vanishingly small compared to other decoherence mechanism, such as
the interaction with the nuclear spins.

In conclusion, we have demonstrated 
all-optical tuning of Raman scattered photons in a single QD. The present experiment opens up the possibility to optically tune
the frequency of photons emitted from two different
QDs into resonance with each other. This could be done for two QDs that are subject to the
same external magnetic field. In addition to the Zeeman shift,
the electron experiences a random Overhauser shift due to hyperfine interaction with
the surrounding nuclei \cite{Khaetskii}. 
The Overhauser shift varies strongly from
one QD to another and consequently the electron ground state splitting 
will in general differ between two QDs even for identical electron
g-factors. In order to resonantly tune two QDs one would need to compensate for this energy difference
by for example using two different driving lasers.
An interesting
follow-up experiment could be the demonstration of two-photon
quantum interference as was done in ions and atoms \cite{Grangier,Monroeions}. This would
constitute a first step towards a probabilistic entanglement scheme.
Beyond its potential for applications in quantum information processing, we envision the
detection of Raman photons as a new tool for studying the
interaction of a single electron spin with the surrounding solid
state.\\
\indent
We thank M. Kroner for helpful discussions and comments on the
manuscript. T.V. acknowledges financial support from the European
Union within the Marie-Curie Training Research Network EMALI.
Furthermore, financial support from the Swiss National Science Foundation
SNSF is gratefully acknowledged.


\begin{thebibliography}{}
\bibitem{linewidth} When the excitation laser is close to resonance $\Delta \leq \Gamma$, we expect a contribution to the signal from scattering due to real excitation into the trion state. However, we cannot resolve this structure due to the resolution of our Fabry-Perot and therefore we fit a simple Lorentzian which gives a reasonable estimate for the linewidth and the number of scattered photons.
\end{thebibliography}
\end{document}